\documentclass{ws-ijmpd}
\usepackage[super,compress]{cite}
\usepackage{epstopdf}
\usepackage{color}
\usepackage{ulem}
\begin{document}
\def\Journal#1#2#3#4{{\it #1} {\bf #2}, (#3) #4}
\def\Book#1#2#3#4{{\it #1} (#2 , #3, #4)}
\def\RPP{{Rep. Prog. Phys}}
\def\PRC{{Phys. Rev. C}}
\def\PR{{Phys. Rep.}}
\def\PLB{{Phys. Lett. B}}
\def\AP{{Ann. Phys (N.Y.)}}
\def\PRD{{Phys. Rev. D}}
\def\ZPA{{Z. Phys. A}}
\def\NPA{{Nucl. Phys. A}} 
\def\JPG{{J. Phys. G }}
\def\PRL{{Phys. Rev. Lett}}
\def\EPJA{{Eur. Phys. J. A}}
\def\NP{{Nucl. Phys}}  
\def\RMP{{Rev. Mod. Phys}}
\def\IJMPE{{Int. J. Mod. Phys. E}}
\def\AJ{{Astrophys. J}}
\def\AJL{{Astrophys. J. Lett}}
\def\AA{{Astron. Astrophys}}
\def\ARAA{{Annu. Rev. Astron. Astrophys}}
\def\MPLA{{Mod. Phys. Lett. A}}
\def\ARNPS{{Annu. Rev. Nuc. Part. Sci}}
\def\LRR{{Living. Rev. Relativity}}
\def\CQG{{Class. Quantum. Grav}}
\def\Journal#1#2#3#4{{\it #1} {\bf #2}, (#3) #4}

\markboth{C. Wibisono, A. Sulaksono}
{Information-Entropic Method: Stability of Stars and Modified Gravity Theories}

%
\catchline{}{}{}{}{}
%

\title{Information-Entropic Method in Studying Stability Bound of Nonrelativistic Polytropic Stars within Modified Gravity Theories}

\author{C. Wibisono}

\address{Departemen Fisika, FMIPA, Universitas Indonesia\\
Depok 16424, Indonesia,\\
catur.wibisono@ui.ac.id}

\author{A. Sulaksono}

\address{Departemen Fisika, FMIPA, Universitas Indonesia\\
Depok 16424, Indonesia,\\
anto.sulaksono@sci.ui.ac.id}

\maketitle

\begin{history}
\received{Day Month Year}
\revised{Day Month Year}
\end{history}

\begin{abstract}
We study the stability of non-relativistic polytropic stars within two modified gravity theories i.e., Beyond Horndeski gravity and Eddington inspired Born Infeld theories using the configuration entropy method. We use spatially localized bounded function of energy density as solutions from stellar effective equations to construct the corresponding configuration entropy. We use the same argument as the one used by the authors of Refs. [1,2] that the stars are stable if there is peak in configuration entropy as a function of adiabatic index curve. Specifically, the boundary between stable and unstable regions which corresponds to Chandrasekhar stability bound is indicated from the existence of the maximum peak while the most stable polytropic stars is indicated by minimum peak in the corresponding curve. We have found that the  value of critical adiabatic indexes of Chandrasekhar stability bound and the most stable polytropic stars predicted by non-relativistic limit of Beyond Horndeski Gravity  and Eddington inspired Born Infeld theories are different to those predicted by general relativity where the corresponding differences depend on the free parameter of both theories.
\end{abstract}


\ccode{PACS numbers:04.50.-h,98.80.-k,11.10.Lm,04.40.Dg}


\section{Introduction}
\label{sec_intro}

In recent years, there has been a lot of interest in studying alternative or modified theories to general relativity (GR) due to 2 reasons (for recent review see Refs. [3-5] and the references therein). The first reason is despite the outstanding successes of GR passed almost all precision tests in intermediate energy scale, GR is incapable to explain satisfactory issues such as  dark matter and dark energy existence. Note that if gravity is governed by GR, substantial amount of dark matter should be existed in galaxies and clusters while dark energy are also required in order the apparent accelerating expansion of the Universe (cosmic expansion) can be explained. It seems that around 96 $\%$ of the Universe should be in the form of these electromagnetically non-interacting matter and energy densities~\cite{CFPS2012}.  The second reason is that the dynamical evolution of matter fields in GR such as Big Bang or those forming in the gravitational collapse of matter field, is disturbed by the formation of singularities. This can be considered as a signal that the theory is breakdown in this limit~\cite{Pani2012}. Note that  according to Ref.[3], Beyond Horndenski gravity (BHG)  and Eddington inspired Born Infeld (EiBI) theories are the ones of the theories which after very stringent sets of requirements can be classified as plausible alternatives to GR in the context of strong field tests. EiBI and  wide class of BHG theories of $G^3$ type yield modification of the Newtonian hydrostatic equilibrium equations. The modifications have impact on properties of polytropic non-relativistic stars including their stability. We need also to note that the structure and cosmological properties of a number of modified gravity theories such as f($\it R$) and Ho\~rava-Lifshitz f($\it R$) gravity, scalar-tensor theory, string-inspired and Gauss-Bonnet theory, non-local gravity, non-minimally coupled models, and power-counting renormalizable gravity are comprehensively discussed and reviewed~\cite{NojiriOdintsov2011}. They have found that some versions of above theories may be consistent with local test of gravity and may provide a qualitatively reasonable unified description of inflation with the dark energy epoch. The gravitational alternative for dark energy and the coincidence problem as a manifestation of the universe expansion in the context of f($\it R$), f($\mathcal{G}$) and f($\it R$,$ \mathcal{G}$) are also reviewed~\cite{NojiriOdintsov2007}. It is also reported recently that peculiarities and some novel physical consequences of compact object predictions in frame of some versions of f($\it R$)~\cite{Arapoglu2011,Astashenok2013,Astashenok2014,Astashenok2017},  f($\mathcal{G}$)~\cite{Astashenok2015}, Scalar-Tensor gravity\cite{Sotani2017}, and minimal dilatonic gravity\cite{Fiziev2017} theories. It is also worth to note that beside BHG and EiBI, the f($\it R$), Scalar-Tensor-Vector gravity, and forth order gravity theories can also introduce modifications to the Poisson equation in Newtonian limit. These bring up also interesting features inside stellar objects~\cite{Banerjee2017}. It means stability study for these theories is also interesting. However, the corresponding studies are already outside of the scope of this work.

Beyond Horndenski theories capable to predict the cosmic expansion without cosmological constant (dark energy). However the theories need screening mechanism known as  Vainshtein mechanism\cite{Vain1972} in order they are still compatible with local test of gravity. This mechanism use nonlinear features of the corresponding theory to decouple the scales such that it  behaves like GR in solar system while modified the cosmology predictions~\cite{ModGravVein, KoySak2015}. It is shown in Ref.[20] that the modification in Newtonian hydrostatic equilibrium in BHG theories is due to the Vainstein mechanism partially broken inside astrophysical bodies (see review for examples in Refs. [19,21] and the references therein). The properties of dwarf stars~\cite{Sak2015A}, hydrogen burning in low mass stars~\cite{Sak2015B}, galaxy clusters~\cite{SWBKN2016}, stars and galaxies~\cite{KoySak2015} are recently used to constrain parameters $\Upsilon_i$ of BHG theory. Parameter $\Upsilon_i$ can be related to the appearing parameters in effective field theory of dark energy (see Refs. [22-24] and the references therein). Note that the covariant quartic Galileon model which admit a stable self-accelerating solution for the background expansion, leads to $\Upsilon$=1/3~(see Ref. [20] and the references therein). 

Eddington inspired Born Infeld (EiBI) theory belong to larger class i.e., Born-Infeld inspired modifications gravity (see review of this class of theories and their applications in astrophysics, black holes, and cosmology in Ref. [25]). The main advantages of Born-Infeld inspired modifications of gravity compared to other modified theories are these theories have shown an extraordinary ability to regularize the gravitational dynamics, leading to non-singular cosmologies and regular black hole spacetimes in very robust manner without resorting to quantum gravity effects~\cite{JHOR2017} (for more details of EiBI theory see also in Refs. [3,6,26-27] and the references therein). EiBI  theory is proposed for the first time by Banados and Ferreira~\cite{BanadosF} where the Palatini or auxiliary field approach is used to treat  the gravitational analog of a nonlinear theory of electrodynamics known as the Born-Infeld theory~\cite{Born:1934gh}. We need to note that  parameter $\kappa$ in EiBI theory can be constrained by using some astrophysical and cosmological data~\cite{Avelino12}, NSs properties~\cite{QISR2016,Harko13,Pani2012,Pani2011} and the Sun properties~\cite{CPLC2012}. The later are indeed analyzed by using the non-relativistic limit form of EiBI theory.

We need to point out here that the structure and dynamics of the stars and collisionless galaxies (cluster) are often mimicked by polytropic equation of states (EOSs). The EOSs are characterized by polytropic index $n$ and the physics of self gravitating polytrope in the Newtonian limit is governed by  Lame-Emden equation (LEE). For examples, in the case of GR,  Newtonian stars with $n \ge 3$ are unstable against radial oscillations, the spatial extent of a polytropic galaxy (cluster) is infinite for $n \ge 5$, and in isothermal sphere model of galaxies the $n \rightarrow \infty$ limit polytropes is indeed considered. In general, the stiffness of polytropic stars decreases with an increase in the polytropic index (see for example Ref. [34] and the references therein). Therefore $n=1,1.5,3,3$ can be used to describe brown dwarf, red dwarf, main-sequence, and white dwarf, respectively~\cite{SKK2016}.  The authors of Ref. [34] studied the perturbation solution of standard LEE.  Furthermore, we need to note that the authors of Ref. [36] have shown that in the case of  EiBI theory, the star matter described by polytropic EOS with adiabatic index $\gamma > 3/2$ yields singular curvature at the surface of the stars. Whether or not this is consequence of polytropic approximation or whether such singularities can be avoided in other ways is currently unclear~\cite{Berti_etal2015,Kim2014}. We need to note that the authors of Ref. [8] by using polytropic EOS and BHG modified LEE has found universal lower bound of  $\Upsilon$ independently of the details of the EOS.  The authors of Refs. [6,26] also studied the spherically symmetric non-relativistic gravitational collapse of dust within non-relativistic limit of EiBI theory. They have found that for any positive value of  $\kappa$ the evolution ended with pressureless stars rather than singularity. Furthermore, they also have found that there is no Chandrasekhar limit in EiBI theory for the mass of a non-relativistic white dwarf for $\kappa>$0.

The compact objects would be stable, if the system's binding energy $E_b$ is negative definite, namely, $E_b=M-Qm$ where $M$ is the mass of compact objects, $Q$ is the conserved number of particles in the objects, and $m$ is the mass of particle. However, this condition does not suffice to explain the stability of the objects. It is necessary to check the stability by giving radial perturbation (oscillation) to the effective equations describing the compact objects \cite{Candra1964,Shapiro}. If the eigenvalue of the perturbed equations is not imaginary, then the systems would be stable under radial perturbation.  Stellar pulsation of modified LEE  within  BHG theory are studied in Ref. [35]. They have found that brown dwarfs and Cepheid are found to be particular sensitive object to test the corresponding theory. We note also that stellar oscillations in a particular type of scalar-tensor theories are studied~\cite{Sak2013} using polytropic EOS. The author found that the GR condition for stellar stability is altered so that the adiabatic index $\gamma =1+1/n$ can fall below $4/3$ before unstable modes appear. It means the stars  within these theories are more stable. On the other hand, the authors of Refs. [6,26] studied the radial perturbation of non-relativistic polytropic stars and they have found that for perturbation method with $\gamma$ = $4/3$, are marginally stable for any polytropic index $n$ and in their case, the model stable if $\kappa >$ 0 and unstable if if $\kappa <$ 0. The authors of Ref. [42] also has shown that the standard results of stellar stability still hold in EiBI theory where for a sequence of stars with the same EOS, the fundamental mode $\omega^2$ passes through zero at central density corresponding to the maximum-mass configuration is similar to the one found in GR. They also found that the fundamental mode is insensitive to the value $\kappa$, while higher order modes depends strongly on $\kappa$.

In this work, we will study stability bound of modified non-relativistic polytropic stars  predicted by BHG and EiBI theories by using information-entropic concept that commonly called configuration entropy~\cite{Gleiser3,Gleiser2}. The  configuration entropy can provide information about the polytrope's stability in weak field regime. Therefore, it can be considered as alternative of well known perturbation method \cite{Candra1964,Shapiro,Sak2013,Pani2012,SKK2016,SLL2012}. This concept uses the  complexity of spatial distribution of a system in momentum space which are generated from the solution of differential equations which describe the system. It is the extension of Shannon-information theory \cite{Betts} but with using continuous probability distributions. Here, one needs to formulate the probability distribution from  mass density profile to compute the corresponding configuration entropy. The previous studies in applying this concept to study astrophysical objects had been done by Gleiser and collaborators \cite{Gleiser3,Gleiser2}. The authors investigated the stability bound of cold white dwarfs, neutron stars with Oppenheimer-Volkoff equation of state, Q-balls, and interacting scalar field of the boson stars. All objects are studied  in Refs. [1,2] within GR framework. The results are compatible with the ones obtained using standard method such as radial perturbation method. For example, the Chandrasekhar mass of white dwarfs is estimated within 3.73 $\%$ of the correct value~\cite{Gleiser3,Gleiser2}.

This paper is organized as follows. In Sec.~\ref{Netmodgrav}, we provide effective equations for describing the hydrostatic equilibrium equations within BHG and EiBI theories and their possible impacts on stability of polytropic stars. Then we briefly reviewed the formulation of configuration entropy before discussing the configuration entropy results predicted by both theories with emphasizing on the relation between stability bound and free parameter of both theories in Sec.~\ref{ConfiEnt}. The conclusion is given in Sec.~\ref{sec_conclu}.

\section{Hydrostatic equation for modified gravity theories}
\label{Netmodgrav}
Stellar equation governing by Newtonian hydrostatics equilibrium of polytropic stars can be written in terms of dimensionless variables or it is usually called LEE~\cite{Shapiro,YCL2017}. LEE has interesting features such as for $n$ =0, 1, and 5 can be solved analytically while for other $n$ can be solved perturbatively or numerically\cite{YCL2017} and the LEE solutions are independent to central mass density $\rho_c$ of the stars. It can be shown easily from LEE that polytropic EOS with  $\gamma_c$ = $4/3$ is the stability bound of polytropic stars.  Here, we will discuss the modified LEEs obtained from non-relativistic limit from BHG and EiBI  theories and the qualitative impact of free parameter of the corresponding theories  on stability bound of polytropic stars.

\subsection{Beyond Horndeski gravity theory}
BHG is one of modified gravity theory which contains screening mechanism if $R<<r_{V}$. Here $r_{V}$ is Veinstein radius and $R$ is the radius of the corresponding astrophysical body. It means that $r_{V}$ defines the transition between screened and unscreened regime inside astrophysical bodies (see detail discussion of Veinstein mechanism in Ref.[22] and the references therein). The equation of hydrostatic equilibrium obtained from non-relativistic limit of BHG theory can be expressed as \cite{Sak2015A,Sak2015B,ModGravVein}
\begin{equation}
\frac{dP}{dr}=-\frac{G_{N}M\rho}{r^2}-\frac{\Upsilon G_{N}\rho}{4}\frac{d^2M}{dr^2},
\label{eq:lane}
\end{equation}
where $\Upsilon$ is dimensionless free parameter of BHG theory and $G_N=\frac{G}{1+5\Upsilon}$. By multiplying Eq.~\eqref{eq:lane} with $\frac{r^2}{\rho}$ and  followed by differentiating the corresponding expression with $r$ and then  supplementing the corresponding expression with standard mass conservation $\frac{dm}{dr}=4\pi\rho r^2$ as well as the polytropic EOS $P=K\rho^{\gamma}$, we can finally obtain
\begin{equation}
\begin{aligned}
\frac{1}{r^2} \frac{d}{dr}\left[\frac{d\rho}{dr}\left\{r^2K\gamma \rho^{\gamma -2} +\frac{\Upsilon}{4}4\pi G_{N}r^4\right\}+\frac{\Upsilon}{4}8\pi G_{N}r^3\rho \right]
=-4\pi G_N \rho .
\end{aligned}
\label{eq:ggg}
\end{equation}
By using dimensionless variables of mass density  $\theta(\chi)$ and radius $\chi$ which are defined respectively as $\rho \equiv \rho_c\theta^{\frac{1}{\gamma -1}}$ and $r \equiv r_c\chi$ where  $r_c^2\equiv\frac{K\gamma}{4\pi G_N \left(\gamma -1\right)}\rho_c^{\gamma -2}$,
then the differential equation in Eq. (\ref{eq:ggg}) can be further simplified into 
\begin{equation}
\begin{aligned}
\frac{1}{\chi^2}\frac{d}{d\chi}\left[\left\{1+\frac{\Upsilon}{4\left(\gamma -1\right)}\chi^2\theta^{\frac{2-\gamma}{\gamma -1}}\right\}\chi^2\frac{d\theta}{d\chi}+\frac{\Upsilon}{2}\chi^3\theta^{\frac{1}{\gamma -1}}\right]
=-\theta^{\frac{1}{\gamma -1}}.
\end{aligned}
\label{eq:5}
\end{equation}
It can be seen that Eq.~\eqref{eq:5} is independent to  $\rho_c$ because $\Upsilon$ is dimensionless free parameter.

We can obtain the profile of dimensionless mass density $\theta(\chi)$ as a function of dimensionless radius $\chi$ in Eq.\eqref{eq:5} numerically by using 4th order Runge-Kutta method. Note that Eq. \eqref{eq:5} also can be called as modified LEE for non-relativistic limit of BHG theory. As an example, the profile of dimensionless mass densities in the case $\gamma=4/3$ for some $\Upsilon$ values of non-relativistic limit of BHG theory are shown in left panel of Fig.~\ref{Fig:ProfileEnergy}. 

\subsection{Eddington inspired Born   Infeld theory}
The non-relativistic limit of EiBI theory is described by the following modified Poisson equation\cite{BanadosF}
\begin{equation}
\nabla^2 \Phi=4\pi G_N\rho+\frac{\kappa}{4}G_N\nabla^2\rho,
\end{equation}
where $\Phi$ is gravitational potential, $G$ is gravitational constant, $\kappa$ is free parameter of EiBI theory in unit $m^2$, and $\rho$ is the mass density. The corresponding spherical symmetry, hydrostatic equilibrium equation can be written as \cite{Pani2012}:
\begin{equation}
\frac{dP}{dr}=-\frac{G_NM}{r^2}\rho - G_N\frac{\kappa}{4}\rho\frac{d\rho}{dr},
\label{eq:peibi}
\end{equation}
By changing Eq.~(\ref{eq:peibi}) into dimensionless form then we arrive to following expression
\begin{equation}
\frac{1}{\chi^2}\frac{d}{d\chi}\left[\chi^2\theta^{\frac{2-\gamma}{\gamma -1}}\frac{d\theta}{d\chi}\left\{\theta^{\frac{\gamma -2}{\gamma -1}} + \frac{\kappa}{16\pi r_c^2(\gamma -1)}\right\}\right]=-\theta^{\frac{1}{\gamma -1}}.
\label{Lane3}
\end{equation}

Eq.~\eqref{Lane3} can be called modified LEE for non-relativistic limit of EiBI theory. However, Eq.~\eqref{Lane3} still depends on non dimensionless variable $r_c$ because $\kappa$ is not dimensionless parameter. It means that one of a nice feature of Newtonian LEE i.e., independent to $\rho_c$ is loss in the case of EiBI theory. However, it is known that from Ref.[6] that stellar solutions would exist if the following condition is satisfied,
\begin{equation}
\kappa G_N > -\left|\kappa_c\right|G_N=-4K\gamma \rho_c^{\gamma -2}.
\label{constraint}
\end{equation}

Note that for $\kappa>0$, the condition stated in Eq.\eqref{constraint} is always satisfy\cite{Pani2012}. In order to obtain profile mass density from Eq.~\eqref{Lane3}, we use the constraint in Eq. \eqref{constraint} to specify the value of $r_c$ which implicitly depends on the mass density at the center of the star $\rho_c$. Here, we take $\rho_c$= $\rm 8\times 10^{17}~kg~m^{-3}$, and the value of $\kappa_c$ is computed from Eq.\eqref{constraint} by substituting polytropic constant $K$ at $\gamma=4/3$, hence the value of $\kappa_c$= 4.53 $\rm \times 10^8~m^2$ is obtained. From Eq. \eqref{constraint}, we can compute the values of other polytropic constants $K$ by specifying the value of the corresponding $\kappa_c$. The profile of dimensionless mass densities in the case of $\gamma=4/3$ for some $\kappa$ values of EiBI theory are shown in right panel of Fig.~\ref{Fig:ProfileEnergy}.  

\subsection{Mass Profile of Modified Gravities}
\begin{figure}
\centerline{\includegraphics[scale=0.6]{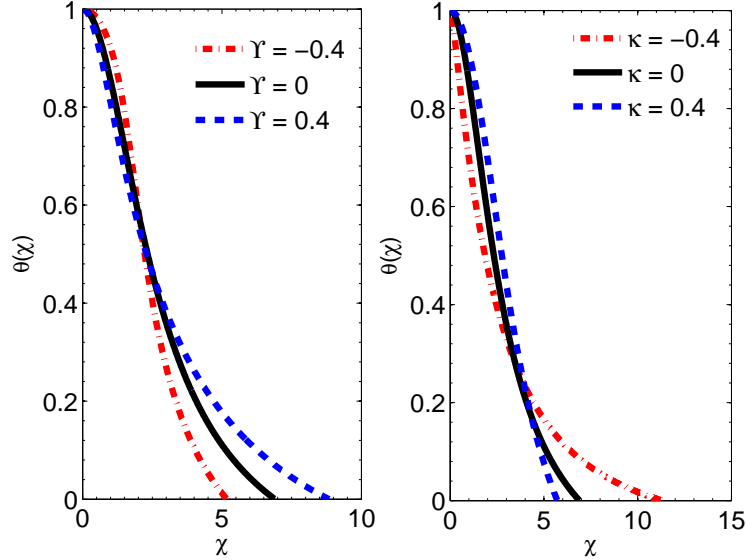}}
\caption{Dimensionless mass density profile for adiabatic index $\gamma=4/3$ for non-relativistic limit of BHG theory (left panel) and non-relativistic limit of EiBI (right panel). The unit of $\kappa = 10^9 $ m$^{2}$.}
\label{Fig:ProfileEnergy}
\end{figure}
It can be seen in Fig.~\ref{Fig:ProfileEnergy} that $\theta(\chi)$ profile for  $\gamma=$4/3 depends on $\Upsilon$ or $\kappa$ and these dependecies are also happen for others $\gamma$. It can be observed that compared to the one of GR ($\Upsilon=\kappa=$0), the difference appears significantly in tail region of the corresponding profiles. In the case of BHG theory,  it is evident that for positive $\Upsilon$, increasing  $\Upsilon$ value  yields larger radius, while for negative $\Upsilon$, increasing  $\Upsilon$ absolute value yields smaller radius. For EiBI theory, the situation is reversed. For positive $\kappa$, increasing $\kappa$ value yields smaller radius, while for negative $\kappa$, increasing $\kappa$ absolute value yields larger radius. These can be understood from Eqs.\eqref{eq:lane} and \eqref{eq:peibi} that the sign and  values of $\Upsilon$ or $\kappa$ would affect the strength of gravity, as a result the radius of the stars would increase or decrease depending on  these parameters. For BHG theory, the result confirms the results obtained in Refs.[22,23].  Note that the difference in  profile of dimensionless mass densities in small $\chi$ of EiBI theory is relative quite pronounced compared to the one of BHG. These differents can be seen from the different in  hydrostatic equilibrium equations of both theories. for EiBI, it depends on $\frac{d \rho}{dr}$ instead of $\frac{d^2M}{dr^2}$. We need to note that one of the interesting feature of EiBI is its compactness ($\frac{GM}{R}$) which increase up to a critical value and then constant when we increase the $\kappa$ value, so that the corresponding compactness never cross the accusal region\cite{Pani2012}. It means that up to critical compactness the increasing of radius is always followed by increasing mass with the same rate. Therefore, in EiBI, increasing $\kappa$ does not always correspond to decreasing the gravity strength.

\subsection{Estimating adiabatic index of stable stars}
\begin{figure}
\centerline{\includegraphics[scale=0.5]{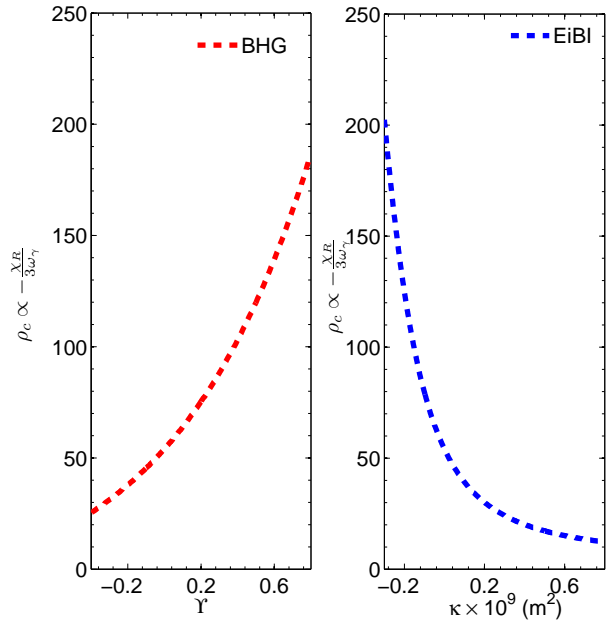}}
\caption{The dependence of central density and variation of modified gravities parameters for $\gamma =4/3$ showing the possibility of stability bound shifting for BHG (left panel) and EiBI (right panel) theories.}
\label{Fig.9}
\end{figure}

In this subsection we will  qualitatively discuss the consequence of dependency of  $\theta(\chi)$ on parameters of both modified gravity theories in stability bound of polytropic stars. It is easy to show that similar to the one of Newtonian~(see for examples Refs. [22,23]),the mass  of polytropic stars of both modified gravity theories  can be also written as
\begin{equation}
M=4\pi\left(\frac{K\gamma}{4\pi G_N (\gamma -1)}\right)^{\frac{3}{2}}\rho_c^{\frac{3}{2}\gamma -2}\omega_{\gamma}(i).
\label{radmass1}
\end{equation}
However, because the dimensionless mass densities of each theory depends on the free parameter of each theory $i$ where, $i=\Upsilon$ for BHG and $i=\kappa$ for EiBI, then in these theories, the  $\omega_\gamma(i)$ i.e.,
\begin{eqnarray}
\omega_\gamma (i)&\equiv& \displaystyle{\int_0^{\chi_R} \chi^2\theta (\chi)^{\frac{1}{\gamma -1}}\, d\chi}\nonumber\\&=& -\chi_R^2\frac{d \theta}{d\chi}|_R, 
\label{radmass2}
\end{eqnarray}
depends also on the free parameter of the corresponding theory. Furthermore it can shown easily (see Refs. [22,23] for the case BHG theory) that for fixed mass and radius of the stars that $\rho_c (i)= -\frac{\chi_R}{3\omega_\gamma (i)} \left[\frac{3M}{4R^3}\right]$.  This relation is explicitly shown in Fig.~\ref{Fig.9} where $\rho_c$ increases by increasing $\Upsilon$ in BHG and $\rho_c$ decreases by increasing $\kappa$ in EiBI theory.
If we assume that the boundary of the real fundamental mode of radial perturbation method is equivalent to $\frac{dM}{d\rho_c}=0$~\cite{Shapiro} applies to all gravity theories then it can be obtained from Eq.~(\ref{radmass1}) that 
\begin{equation}
\frac{dM}{d\rho_c} \sim  \Bigg[\frac{3}{2} \gamma_c-2+\frac{\rho_c}{\omega_\gamma(i)}\frac{d\omega_\gamma(i)}{d \rho_c} \Bigg]=0,
\label{radmass3}
\end{equation}
where the last term in Eq.~(\ref{radmass3}) is zero in the case of GR. This means that in EiBI and BHG the  stability bound of polytropic stars is shifted from stability bound adiabatic index of GR i.e., $\gamma_c$=4/3 to larger or smaller values depending on the value of the corresponding free parameter.

Even, it is not always sufficient quantitatively, the boundary of stability can be also seen from binding energy of the star equal to zero. We can use this for other argument. In standard non-relativistic limit of GR (Newtonian), one arrives at critical adiabatic index for Chandrasekhar limit ($\gamma_c=\frac{4}{3}$). Here, we try to reformulate binding energy $E$ which consist of internal energy which comes from energy of the matter  and gravitational potential energy of the star for modified gravity theories. By using the $1^{st}$ law of thermodynamics and assuming that the process is adiabatic (i.e the entropy across the polytropes are same) and using polytropic EOS, the internal energy $U$ of polytropes can be written as \cite{Shapiro}
\begin{equation}
U=\frac{1}{\gamma -1}\displaystyle{\int_0^R P4\pi r^2\, dr}.
\label{IntEn}
\end{equation} 

The gravitational potential energy $W$ can be obtained from the integral of an element mass $\delta m$ under gravitational potential $\Phi(\textbf{x})$ at point $\textbf{x}$ where it is generally written as \cite{Binney}:
\begin{equation}
\delta W=\displaystyle{\int \delta \rho~\Phi (\textbf{x})\, d^3\textbf{x}}.
\label{GravTerm}
\end{equation}
In our case, the stars have spherically symmetric form, and gravitational potential has relation to density through modified Poisson equation for each modified gravity. However, different to that of Newtonian in the cases of modified gravities, it is difficult to obtained the exact analytical form of gravitational energy from Eq.~(\ref{GravTerm}). In the following we will discuss the approximate form of gravitational energy of BHG and EiBI  modified gravity theories, respectively. 
\subsection{Approximate Gravitational Potential Energy of Modified Gravities}
Modified Poisson equation of BHG could be constructed from its temporal metric \cite{KoySak2015} and in spherical symmetric system, it could be written as follow:
\begin{equation}
\nabla^2 \zeta(r)=4\pi G_N \rho,
\label{PoissonCG1}
\end{equation}
where for BHG theory, we define $\Phi(r)$ as
\begin{equation}
\Phi(r) \equiv \zeta(r)+\frac{\Upsilon G_N}{4}4\pi r^2\rho(r).
\label{PoissonCG2}
\end{equation}
It means that $\zeta(r)=\Phi_{std}(r)$ has similar structure to standard Newtonian potential. To simplify our problem, we assume that every point in gravitation body feels the same average potential coming from the contribution of second term (BHG correction) of Eq. \eqref{PoissonCG2}. Therefore, in this mean field approximation that the variation of this correction term is vanish ($ \delta \Big[r^2\rho(r)\Big]\approx $0). In consequence, we could apply $\delta \zeta\approx \delta \Phi$. Then by using Eq.\eqref{GravTerm} and Eq.\eqref{PoissonCG1}, the variational of gravitational potential energy of BHG can be written as:
\begin{equation}
\begin{aligned}
\delta W = \frac{1}{4\pi G_N} \displaystyle{\int \Phi (r)\delta~\Big[\nabla^2 \zeta(r)\Big] \, d^3r} 
\approx \delta\Bigg[\frac{1}{8\pi G_N}\displaystyle{\int \Phi(r)\nabla^2 \Phi(r)\, d^3r}\Bigg].
\label{GravTerm2X}
\end{aligned}
\end{equation}
Therefore,
\begin{equation}
W\approx \frac{1}{8\pi G_N}\displaystyle{\int \Phi(r)\nabla^2 \Phi(r)\, d^3r}.
\label{GravTerm2}
\end{equation}
By Evaluating Eq. \eqref{GravTerm2} by using Eq. \eqref{PoissonCG2}, and neglecting the term involving second order of $\Upsilon$ and the one with the term involving $\frac{d^2\rho}{dr^2}$, thus equation above becomes:

\begin{equation}
W \approx \frac{1}{2}\displaystyle{\int \left[\zeta (r)\rho +\frac{\Upsilon}{4} \left\{r^2\rho^2 4\pi G_N +6\zeta(r)\left(\rho +r\frac{d\rho}{dr}\right)\right\}\right]\, d^3r},
\end{equation} 
and taking the explicit function of $\zeta(r)$ as solution of Eq. \eqref{PoissonCG1}, then transforming the first term of $W$ into Chandrasekhar potential energy tensor \cite{Binney}, we can obtain: 

\begin{equation}
\begin{aligned}
W &\approx \displaystyle{\int_0^R -\frac{G_Nm(r)}{r}4\pi r^2 \rho(r)\, dr}
+\frac{1}{2}\frac{\Upsilon}{4}(4\pi)^2 G_N\displaystyle{\int_0^R r^4\rho^2(r)\, dr}\\
&+\frac{3\Upsilon}{4}\displaystyle{\int_0^R -\frac{G_Nm(r)}{r}\left[\rho(r)+r\frac{d\rho(r)}{dr}\right]4\pi r^2\, dr}.
\end{aligned}
\label{GravTermCG}
\end{equation}

By adding Eq. \eqref{IntEn} and Eq. \eqref{GravTermCG}, the analytic form of approximate binding energy $E$ of polytropes within BHG theory can be obtained.  

Modified Poisson equation of non-relativistic limit of EiBI can be also expressed as follow \cite{Pani2012, Pani2011}:
\begin{equation}
\nabla^2 \eta(r)=4\pi G\rho,
\end{equation}
where for EiBI theory, we define $\eta(r)$ as
\begin{equation}
\eta (r) \equiv \Phi(r)-\frac{\kappa G}{4}\rho(r).
\label{GravNREiBI}
\end{equation}
By using the similar arguments with the ones used in deriving gravitational potential energy of BHG,  
we can obtain:
\begin{equation}
\begin{aligned}
W &\approx \displaystyle{\int_0^R -\frac{G_Nm(r)}{r}4\pi r^2 \rho(r)\, dr}
+\frac{1}{2}\frac{\kappa}{4}(4\pi) G_N\displaystyle{\int_0^R r^2\rho^2(r)\, dr}\\
&+\frac{\kappa}{4}\displaystyle{\int_0^R -\frac{G_Nm(r)}{r}\left[r\frac{d\rho(r)}{dr}\right]\, dr}.
\end{aligned}
\label{GravTermEiBI}
\end{equation}
It can be seen that the second and the third terms in Eq.~(\ref{GravTermEiBI}) can be shifted the critical adiabatic index $\gamma_c$ into values larger or smaller than $\frac{4}{3}$ depending on the $\kappa$ value used. 

\begin{figure}
\centerline{\includegraphics[scale=0.6]{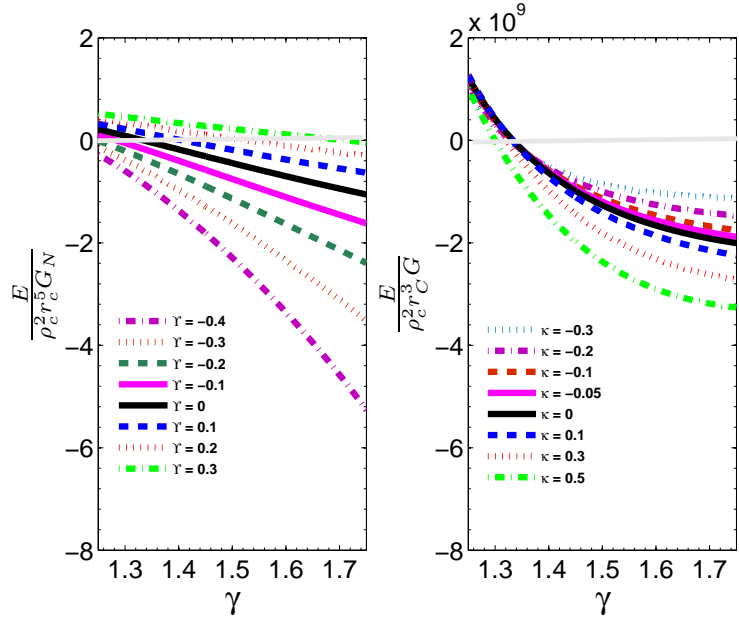}}
\caption{Binding energy of polytropes in BHG theory (left panel) and EiBI (right panel). The unit of $\kappa$ is 10$^9$ m$^2$.}
\label{FigureX1}
\end{figure}

The binding energy as a function of $\gamma$ of both modified gravities can be seen in Fig.~\ref{FigureX1}. It can be seen from the left panel of Fig.~\ref{FigureX1} that due to the role of second and the third terms in Eq.~(\ref{GravTermCG}), the critical adiabatic index  can be shifted  from $\gamma_c=\frac{4}{3}$ depending on the value of  $\Upsilon$. It can be shown that decreasing the values of $\Upsilon$ is always followed by the shifting of critical adiabatic index to the smaller value of $\gamma_c$. Therefore, for smaller value $\Upsilon$, the range of  $\gamma$ of stable stars is  wider. From the right panel of  Fig.~\ref{FigureX1}, it can be observed that in the case of negative values of $\kappa$ the change of critical adiabatic index which indicates by zero values of binding energy $E$ do not appear significantly. Whereas, in the case of positive values of $\kappa$, the increasing values of $\kappa$ is followed by increasing the critical adiabatic index. It means different to those of BHG theory, for larger but positive value of $\kappa$, the range of  $\gamma$ of stable stars is  wider.

In the next section, we discuss in more quantitative way the influence of free parameter of BHG and EiBI modified gravity theories to the stability of polytropic stars  using configuration entropy method instead of radial perturbative method. We also compare the configuration entropy results to the ones obtained by observing the zero of approximate binding energies obtained in this section.

\section{Configuration entropy and stability bound}
\label{ConfiEnt}

\subsection{Formalism}
Here, we will first briefly review the configuration entropy formalism that related with stability bound proposed by the authors of Refs. [1,2]. For any spatially localized bounded functions $f(\textbf{x})$, and their Fourier transforms in momentum space $F(\textbf{k})$ forms, we can have following relation:
\begin{equation}
\displaystyle{\int_{-\infty}^{\infty} \left|f(\textbf{x})\right|^2\, d^dx}=\displaystyle{\int_{-\infty}^{\infty} \left|F(\textbf{k})\right|^2\, d^dk}.
\end{equation}
Then we can define the modal fraction $f(\textbf{k})$ as \cite{Gleiser1}
\begin{equation}
f(\textbf{k})=\frac{\left|F(\textbf{k})\right|^2}{\displaystyle{\int \ \left|F(\textbf{k})\right|^2\, d^dk}},
\label{eq:ModFrac}
\end{equation}
where the integration in Eq. (\ref{eq:ModFrac}) is evaluated over the possible range of $\textbf{k}$, while $d$ states the number of spatial dimensions. In analogy with Shannon information theory \cite{Betts} which states that, $S_s=-\displaystyle{\sum p_i \log p_i}$. For general non-periodic functions in the continuous distribution, the configuration entropy can be written as \cite{Gleiser1}
\begin{equation}
S_c[f]=-\displaystyle{\int \ \tilde{f}(\textbf{k})\ln[\tilde{f}(\textbf{k})]\, d^dk},
\end{equation}
where $\tilde{f}(\textbf{k})=f(\textbf{k})/f(\textbf{k})_{max}$ is normalized modal fraction and $f(\textbf{k})_{max}$ is maximum fraction which is given by the longest physical mode of the system, $\left|k_{min}\right|=\pi /R$, or by zero mode $\textbf{k}=0$. Following Refs. [1,2], here we use energy density $\rho(\textbf{r})=\rho(r)$ of spherically symmetric stellar equations as the spatially localized bounded function. Hence the Fourier transform of energy density profile can be explicitly written as \cite{Gleiser2}
\begin{equation}
F(\textbf{k})=4\pi\displaystyle{\int_0^R \frac{\rho (r)r\sin(kr)}{k}\, dr}.
\end{equation}
The modal fraction and configuration entropy can be expressed in dimensionless variables of energy density and radius. We use the following standard LEE scaling such as $r=r_c\chi$ and $\rho=\rho_c\theta(\chi)^{\frac{1}{\gamma -1}}$ for this purpose. Therefore, the normalized modal fraction can be written as:
\begin{equation}
\tilde{f}(k)=\frac{\left|F(r_ck)\right|^2}{\left|F(r_ck)_{max}\right|^2}=\frac{\left|F(r_ck)\right|^2}{\left|F(\frac{\pi}{\chi_R})\right|^2},
\label{Modfrac1}
\end{equation}
where the Fourier transform of energy density in Eq.\eqref{Modfrac1} can be expressed as
\begin{equation}
F(r_ck)=4\pi \rho_c r_c^{3}\displaystyle{\int_0^{\chi_R} \frac{\theta(\chi)^{1/(\gamma -1)} \chi \sin(r_ck\chi)}{r_ck}\, d\chi }.
\label{Modfrac2}
\end{equation}
Therefore, the configuration entropy $S$ can be expressed as following dimensionless form
\begin{equation}
\begin{aligned}
S=-4\pi r_c^{-3} \displaystyle{\int_{\alpha_{min}}^{\infty} \frac{\left|F(\alpha)\right|^2}{\left|F(\frac{\pi}{\chi_R})\right|^2} \log\left(\frac{\left|F(\alpha)\right|^2}{\left|F(\frac{\pi}{\chi_R})\right|^2}\right)\alpha^2\, d\alpha},
\end{aligned}
\label{CEntro}
\end{equation}
where $\alpha \equiv r_ck$, so that $\alpha_{min}=\pi/\chi_R$. We will study the bound of the stars stability numerically using Eq.~\eqref{CEntro} in the next subsection.

\subsection{Results and Discussions}

Here we will discuss the  numerical configuration entropy results predicted by non-relativistic limit of BHG and EiBI theories.

\begin{figure}
\centerline{\includegraphics[scale=0.6]{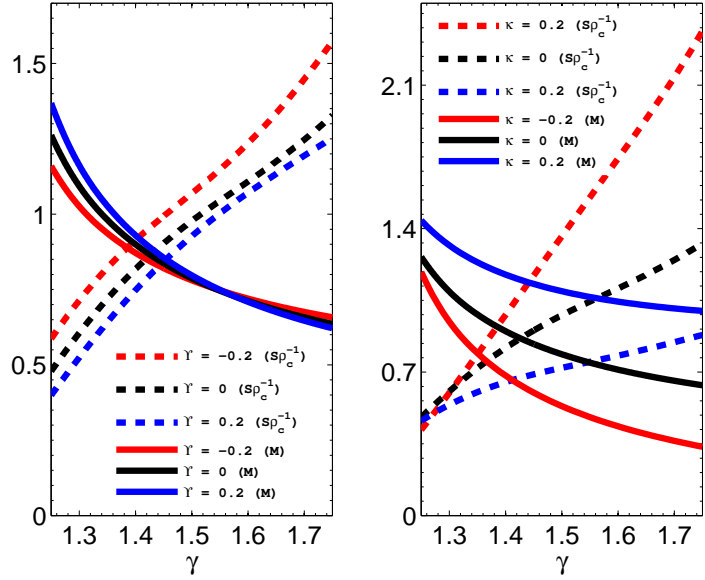}}
\caption{Configuration entropy and mass for non-relativistic limit of BHG theory as a function of $\gamma$ for several values of $\Upsilon$ (left panel) and for non-relativistic limit of EiBI theory as function of $\gamma$ for several values of $\kappa$ (right panel).}
\label{Fig.5}
\end{figure}

Inverse scaling between configuration entropy and mass concept is crucial in this method because the existence of this scaling ensures that the existence of a saddle ridge at exactly the same $\gamma_c$ value in contour plots for mass as a function of central density and $\gamma$ and for configuration entropy as a function of central density and $\gamma$. Please see detail discussion about this scaling in Refs. [1,2]. In Figs.~\ref{Fig.5}, we show the relation between configuration entropy $S\rho_c^{-1}$ in the unit of $\left(\frac{K}{4\pi G}\right)^{\frac{-3}{2}}\rho_c^{2-3/2\gamma}$ and mass $M$ in the unit of $\left(\frac{K}{4\pi G}\right)^{\frac{3}{2}}200\rho_c^{3/2\gamma -2}$ from non-relativistic limit of BHG and EiBI theories, respectively for several values of $\kappa$ and $\Upsilon$. The analytical formulation regarding this scaling is similar to that used in Ref.[1,2]. This is due to the fact that the formulation of mass of the polytropes and configuration entropy of non-relativistic limit of BHG and EiBI theories are similar to that of Newtonian formulation. It can be seen that in general, the inverse scaling relation between the mass of polytropes and their configuration entropy in GR\cite{Gleiser3,Gleiser2} are still existed in non-relativistic limit of BHG and EiBI theories. However, by comparing the scaling results of both theories with the one of GR in detail, it can be seen that the scaling trend predicted by GR is more closer to that of BHG theory than that of EiBI theory, specially in the region of small $\gamma$. Eventually, these results confirm us that the method used in Refs. [1,2] is still valid to investigate the stars stability within non-relativistic limit of BHG and EiBI theories.

\begin{figure}
\centerline{\includegraphics[scale=0.6]{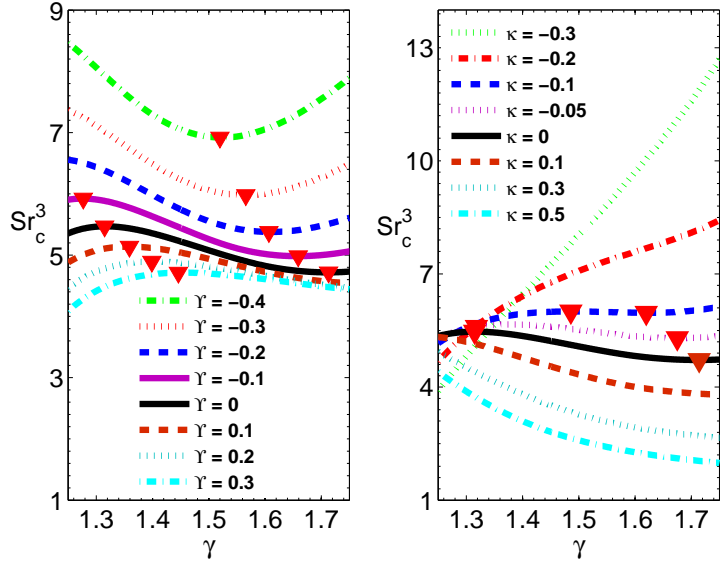}}
\caption{Configuration entropy as a function of $\gamma$ of non-relativistic limit of BHG theory (left panel) and Configuration entropy as a function of $\gamma$ of non-relativistic limit of EiBI (right panel)}
\label{Fig.7}
\end{figure}

In Figs.~\ref{Fig.7}, we plot configuration entropy (CE)  vs $\gamma$ for  non-relativistic limit of BHG and EiBI theories for both positive and negative values of each parameter of the corresponding theory. Here CE is presented by  $Sr_c^3$. The curves with black line are the configuration entropy of Newtonian polytrope. In the case of Newtonian polytrope, we reproduce here, the result obtained in Refs.[1,2]. It can be seen that the Chandrasekhar limit with $\gamma_c=4/3$ lies near the maximum of CE with difference around $1.1\%$ . Note that $ \gamma=4/3$ is adiabatic index for ultra-relativistic EOS. According to Chandrasekhar formulation, the stable stars in Newtonian gravitation lies in $\gamma > 4/3$. Meanwhile, the lower limit of CE lies around $3\%$ close to that of $\gamma_c =5/3$. According to Ref. [2], star with this polytropic EOS is the most stable Newtonian star. We extend this analysis to BHG and EiBI theories, where we connect the stability of polytropes by observing the peak of CE with respect to the variation of the corresponding parameter of each theory. The maximum peak gives boundary between unstable and stable regions and the minimum peak corresponds to the adiabatic index of the most stable polytropic star. 

For non-relativistic limit of BHG theory, we displayed CE profile for several values $\Upsilon$. We obtain that for positive values of $\Upsilon$ the lower bound of stability which is indicated by the  local maximum of CE, shift to the larger value of $\gamma$ if the $\Upsilon$ value increases. While, for negative values of $\Upsilon$, the local maximum CE shift to the smaller value of $\gamma_c$ up to  $\gamma=0$ when the absolute value of  $\Upsilon$ increases. If we see these results from perspective scaling relation in Fig.~\ref{Fig.5}, it is obvious that if we compared to the one of Newtonian, the Chandrasekhar mass or maximum mass would be decreased or increased depending on the value and sign of $\Upsilon$. For $\Upsilon <-0.1$ there are no maximum local of CE observed, and we argue that there are no Chandrasekhar limit for $\Upsilon <-0.1$ where adiabatic index $\gamma$ range runs from 1.25 to 1.75. It can be also observed that  the minimum peak corresponds to the adiabatic index $\gamma_c$ of the most stable polytropic star can not be existed for positive value of $\Upsilon$ while for negative  value of $\Upsilon$, compared to that of GR with $\gamma_c =5/3$, they are shifted to larger value of $\gamma_c$ depending on the absolute value of $\Upsilon$ used. Note that for example, the main sequence stars which are well modeled by $\gamma =4/3$, in BHG theory, these stars are more suitable described using $\Upsilon \leq 0$. Furthermore, we unable to obtain the converge results for dimensionless  energy densities profiles for $\Upsilon < -0.6$ because in this range of $\Upsilon$ the $\frac{dp}{dr}$ becomes negative. Our results agree with Refs. [7,18,24] which states that in order to have stable spherically symetic configuration one need to restrict $\Upsilon > -2/3$. This result ensures us that CE can still be applied for this case. In addition for $\Upsilon \geq 0.7$, the configuration entropy curve with respect to adiabatic index always increases which indicates instability region since there is no maximum local of corresponding curve. Therefore, we conclude that according to CE method the stable polytropic stars are only exist within the range of $-0.6 \leq \Upsilon < 0.7$ . The later is quite consistent the $\Upsilon$ constraint obtained by the author of Ref. [22]. Note that also for the case of White Dwarfs the the stable configurations is found in more restricted range i.e.,  $-0.18 \leq \Upsilon < 0.27$.

In non-relativistic limit of EiBI theory,  the existence of stellar solution determined by constraint $\kappa \geq \left|\kappa_c\right|$ does not imply that the polytopic stars are always stable. We can study the stability of the theory by using the same analysis as the ones used in BHG theory i.e., observing CE profile of several values of $\kappa$. The CE trends obtained by using this theory are rather different compared to those of BHG theory due the difference in modal fraction trends of both theories. We have found for this theory that polytropic stars would be stable for all positive values of $\kappa$. There are also no Chandrasekhar limit for $\kappa>0$, since there are no peak of the CE observed. We argue that for positive values of $\kappa$, the stars are always stable if the constraint  $\kappa \geq \left|\kappa_c\right|$ is satisfy. For negative value of $\kappa$, particularly for $\kappa \leq -0.1$ we do not find  a peak and the slope of CE increases by increasing $\gamma$. Therefore, we conclude that for $\kappa \leq -0.1$ the polytropic stars are unstable. Our results slightly different from the results of Ref. [6,16], which stated that there are no Chandrasekhar limit for any positive values of $\kappa$ and the stars are unstable for  $\kappa \leq 0$. Furthermore, we also have found  no minimum CE in our used range of $\gamma$ for $\kappa \geq -0.1$. It means that the most stable polytropic stars like polytropic star with $\gamma_c=5/3$ in GR are not observed in non-relativistic limit of EiBI theory.

\begin{figure}
\centerline{\includegraphics[scale=0.6]{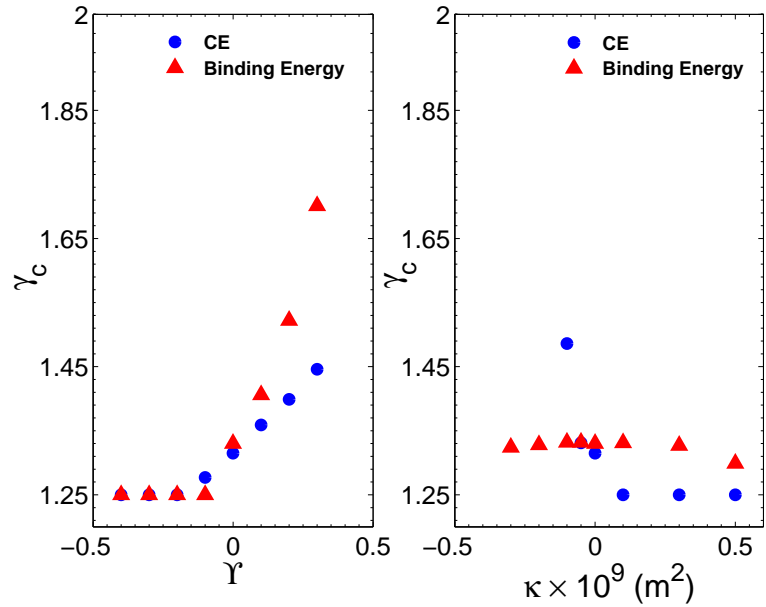}}
\caption{Critical adiabatic index $\gamma_c$ plotted against free parameters of each modified gravity (BHG or EiBI). Both theories, are calculated using configuration entropy and binding energy.} 
\label{Fig.10}
\end{figure}

In Fig.~\ref{Fig.10} we compare the critical adiabatic index $\gamma_c$ as a function of $\Upsilon$ for both modified theories of gravities i.e., BHG and EiBI where they are obtained by using CE method and approximate binding energy method. Our results indicate that the trends of shifting peak of CE as stability indicators which are predicted by both modified gravity theory, are quite consistent to the ones predicted by observing zero value of approximate binding energy. Specially for BHG theory the similarity is quite pronounced where for $\Upsilon \le$ -0.1 both methods predict constant value of $\gamma_c$ but for  $\Upsilon >$ -0.1, the trends of  $\gamma_c$ predicted by both methods increase linearly. On the other hand for EiBI theory for $\kappa >$ 0 the trend of $\gamma_c$ predicted by CE is constant while the ones of binding energy tend also to be constant. For $\kappa =$ -0.05, the predictions of both method  quite compatible but for  $\kappa =$ -0.1, the $\gamma_c$ difference predicted by both methods is relative large. Note that different to that of CE method, binding energy method predicts  $\gamma_c$ values for $\kappa \le$ -0.2 which are known as unstable according the  Pani {\it et. al} results~\cite{Pani2012}. We need to emphasize here the small numerical difference in the predictions for both methods is fairly reasonable, because the critical stability bound obtained through binding energy is rather rough due to some approximations used in deriving gravitational potential energy of both modified theories.

In this end, our findings indicate that the CE method is also quite reasonable to determine the stability not only for GR but also  for  BHG and EiBI modified gravity theories.  Of course the ultimate justification for applicability of this method for modified gravity theories is by comparing the CE method results with the ones predicted by radial perturbation method. However, this is already in outside of the scope of our work. We leave it  for our next future work.

\section{Conclusions}
\label{sec_conclu}
We study stability  of non-relativistic polytropic stars within BHG and EiBI theories by using configuration entropy (CE) method. The results are compared to the one of GR\cite{Gleiser3,Gleiser2}. We have shown that in general the behavior of stability  of both theories are rather different compared to that of GR. The Chandrasekhar limit and the status of the existence of the most stable polytropic stars of EiBI and BHG theories are not the same to ones obtained by using Newtonian gravity and the corresponding prediction differences depend on free parameter of theories. For BHG theory, we obtain that for $\Upsilon < -0.1$  the Chandrasekhar limit is not exist while for  $\Upsilon >0 $,  the corresponding $\gamma_c$ of the Chandrasekhar limit of BHG theory is shifted to larger  than  $\gamma_c$ = 4/3. In consequence that the maximum mass of the polytrope star of BHG theory is smallar than that of GR. We also argue that the value of $\Upsilon$ which gives stable polytrope stars are in the range $-0.6 \leq \Upsilon < 0.7$. For EiBI theory, we have shown that for $\kappa \geq \left|\kappa_c\right|$ stellar solution would always exist, and the stable polytropic star exists if the values of $\kappa \geq -0.1$. Furthermore, in our used range of $\gamma$ for $\kappa \geq -0.1$ the most stable polytropic stars like the polytropic star with $\gamma_c=5/3$ in GR  are not observed.

\section*{ACKNOWLEDGMENT}
CW thanks to I. Prasetyo for discussions and to A. I. Qauli for helping in numerical computation. We are partially supported by the UI's PITTA grant No.~622/UN2.R3.1/HKP.05.00/2017.


\end{document}